\documentclass[a4paper,10pt]{article}
\usepackage{graphicx,amsmath,natbib}
\def\sn2001el{SN\,2001el~}
\def\asn1993{SN\,1993J~}
\def\bsn2005{SN\,2005cf}
\def\csn2004{SN\,2004A}
\def\dsn2001ay{SN\,2001ay~}
\def\grb{GRB 050814}
\def\aap{A\&A\,  }
\def\aj{AJ  }
\def\apj{ApJ\,  }
\def\apjl{ApJ\,  }

\def\mnras{MNRAS\,  }

\def\nat{Nature\,  }

\title
{
Relativistic scaling laws
for the  light curve in  supernovae
}
\author{L. Zaninetti   \\
Dipartimento di Fisica \\
Via Pietro Giuria 1    \\
10125, Turin, Italy    \\
\footnote{zaninetti@ph.unito.it}\hspace{0.08cm}
{Corresponding author: zaninetti@ph.unito.it}}
\begin{document}
\maketitle
\begin{abstract}
In order to explain light curve (LC) for
Supernova  (SN) we derive a classical formula for
the conversion of the flux of kinetic energy into radiation.
We then introduce a correction for the absorption
adopting  an optical depth as function of the time.
The developed framework allows to fit the LC   of
type Ia  SN 2005cf ( B and V )   and  type IIp  SN 2004A
(B,V,I and R ).
A relativistic formula for the flux of kinetic energy is also derived
in terms of a Taylor expansion and the application is done to the LC of
GRB 050814.
The decay of the radioactive  isotopes as a driver the LC
for SNs is also reviewed and a new
formulation  is  introduced.
The  Arnett's  formula for
bolometric luminosity is corrected
for the optical depth and applied to SN 2001ay.
\end{abstract}

\section{Introduction}

The term light curve (LC)  for  Supernova  (SN)
usually denotes  the behavior  of the
apparent/absolute  visual magnitude as function of the time.
The development   of the multiwavelength astronomy
fixes  the  wavelength passband, i.e BVRI,
or the frequency  $\nu$, i.e. 15.2 GHz,
or the  energy, i.e. 1kev.
Further on
in  gamma, X and
radio astronomies
the flux or  the count rate  are
recorded  rather than the magnitude, see as an example
\cite{Fong2012}.
The first model to be considered  is  connected
with  the radioactive decay
\begin{equation}
L = L_{\lambda,0} \exp (- \frac{t}{\tau_n}) \quad ,
\end{equation}
where $L$ and $L_{\lambda,0}$ are the
luminosities  at time $t$ and
at $t=0$ respectively,
$\lambda$  is the considered wavelength
and $\tau_n$ is the typical lifetime, see
\cite{deeming}.
On introducing the apparent magnitude
$m_{\lambda}$, the previous formula becomes
\begin{equation}
m_{\lambda} = k^{\prime}_{\lambda}  +1.0857 (\frac{t}{\tau_n})
\quad ,
\label{mstandard}
\end{equation}
where $k^{\prime}_{\lambda}$  is a constant.
The most important
radioactive isotopes are  $^{56}$Ni with
$\tau_n=8.757\, d$
and  $^{56}$Co with $\tau_n = 111.47 \,d$.
The analysis  of many authors has
shown that the  decay  of one of these two radioactive
isotopes fit  only few days of a typical LC,
see  \cite{Smith2007}.
At the same time  the spectral index in the radio
of \asn1993
is constant after  700 days, see Figure 8 in
\cite{MartiVidalb2011} and this observational fact
points  toward the presence of synchrotron emission
having  flux , $F(\nu) \propto  \nu^{-\alpha}$
with $\alpha \approx 0.7$.
The hypothesis of the synchrotron emission  in SNs
is not new and we now report some applications among others:
GRBs, see \cite{Preece2002,Beniamini2013,Burgess2014}
and  Supernovae Remnants  (SNRs), see
\cite{Katsuda2010,Miceli2013}.
The presence of the synchrotron emission makes attractive
the analysis  of a turbulent cascade
from the large scale
to the small scales where
presumably the relativistic electrons are accelerated.
Insofar we have isolated two completely different physical
mechanism for the source of radiation in the LC:
(i) the number of radioactive isotopes as function
of the time,
(ii)
the flux  of mechanical kinetic energy which
is the driver for the power injected in the turbulent
cascade.
The fact that the spectral index in the optical regime
varies  considerably
with the time points toward a variable
optical thickness  as function of the time and of the
considered pass-band.
The basic idea is that the optical thickness is low
at the beginning of the LC and it increases its value
with time.
A series of questions can now be posed
\begin{itemize}
\item
Can we build a formula for the flux/magnitude versus time
relationship in the framework of conversion
of the mechanical luminosity into radiation?
\item
Can we introduce the correction for variable optical
thickness introducing a dependence of the optical thickness
with time?
\item
Can the new developed framework be applied to the various LCs
which arise from the various typologies of LCs such as
type Ia, Ib, Ic or type IIb, II-L, II-p, IIn?
\item
Can the radioactive model in it's various versions model
the most common LCs?
\end{itemize}

\section{Preliminaries}

Here we first  introduce an elementary  equation
of motion and then we assume  a linear relationship
between the mechanical and the observed luminosity
at a given frequency  $\nu$.

\subsection{The simplest equation of motion}

The equation of   the expansion  of a  SN
in the first ten years
can  be modeled by a power law  of the type
\begin{equation}
R(t) = R_0 (\frac{t}{t_0})^{\alpha}
\label{rpower}
\quad ,
\end{equation}
where
$R$ is the radius  of the expansion,
$t$ is the time,
$R_0$ is the radius  at  $t=t_0$
and  $\alpha$ is an exponent which  can be found from a
numerical analysis.
The velocity is
\begin{equation}
V(t) = \alpha R_0 (\frac{1}{t_0})^{\alpha} t^{(\alpha-1)}
\quad .
\label {vpower}
\end{equation}
As   an example  in the case  of  \asn1993 we have
$\alpha$ = 0.828.
The   rate of transfer of
mechanical energy, $L_{m}$ is
\begin{equation}
L_{m}(t) = \frac{1}{2}\rho (t)4 \pi R(t)^2 V(t)^3
\quad .
\end{equation}
We now  assume  that the density in front of the
advancing  expansion scale as
\begin{equation}
\rho(t) = \rho_0 ( \frac{R_0}{R} ) ^{d}
\quad ,
\end{equation}
where $d$ is a parameter  which allows
to match the observations;
this  assumption is not new and as an example \cite{Nagy2014}
quotes d=3.
The    mechanical luminosity  for the power law dependence of the
radius becomes
\begin{equation}
L_{m}(t) = L_0 (\frac{t}{t_0})^{5\alpha -d \alpha -3}
\quad ,
\label{lmalpha}
\end{equation}
where $L_0$ is the luminosity at $t=t_0$.

\subsection{The emitted radiation}
\label{sectransition}

The energy  fraction  of
the mechanical luminosity, $L_{\nu}$,
deposited
in the frequency $\nu$ is assumed to be proportional
to the mechanical luminosity through a constant
$\epsilon_{\nu}$
\begin{equation}
L_{\nu} = \epsilon_{\nu}  L_m
\quad .
\end{equation}
The flux  at  frequency  $\nu$ and  distance  D
is
\begin{equation}
F_{\nu} = \frac{\epsilon_{\nu}  L_m}{4 \pi D^2}
\quad .
\label{flux_classical}
\end{equation}
The problem of  the absorption
can be parametrized
introducing a
slab of optical thickness $\tau_{nu}$.
The emergent  intensity $I_{\nu}$ after
the entire slab is
\begin{equation}
I_{\nu} = \int_0^{\tau_{\nu}} S_{\nu} e^{-t} dt
\quad ,
\end{equation}
where  $S_{\nu}$ is a uniform source function.
The integration gives
\begin{equation}
I_{\nu} = S_{\nu} (1-e^{-\tau_{\nu}})
\quad ,
\end{equation}
see formula 1.30 in \cite{rybicki}.
A {\it first} model for the
optical thickness assumes a power law dependence
\begin{equation}
\tau_{\nu} = a_{1,\nu} t^{a2,\nu}
\quad ,
\label{taupower}
\end {equation}
where $a_{1,\nu}$ and  $a_{2,\nu}$
are two coefficients  which can be found
from the astronomical
data.
The flux corrected for absorption in
the power law case, $F_{\nu}$, is
\begin{equation}
F_{\nu,c}=
\frac{\epsilon_{\nu}  L_m}{4 \pi D^2}(1-e^{-\tau_{\nu}})
\quad .
\end{equation}
An expression for  the flux  of the {\it first} model
can be obtained
inserting the simplest equation of motion for the $R-t$
 dependence
in the mechanical luminosity as given by (\ref{lmalpha})
\begin{equation}
F_{\nu,c} =
F_{\nu,0}(\frac{t}{t_0})^{5\alpha -d \alpha -3}   (1-e^{-\tau_{\nu}})
\quad ,
\label{fluxfirst}
\end{equation}
where $F_{\nu,0}$ is the flux a $t=t_0$.
This formula is useful  when we have  the flux versus the time,
as an example  Jansky versus JD.
The  absolute/apparent magnitude version of
for the {\it first}  model  is
\begin{equation}
m(t)  = \frac
{
2.5\,\ln  \left( t \right) \alpha\,d- 12.5\,\alpha\,\ln  \left( t
 \right) + 7.5\,\ln  \left( t \right) - 2.5\,\ln  \left( 1-{{\rm e}^{-
a_{{1}}{t}^{a_{{2}}}}} \right)
}
{
\ln  \left( 2 \right) +\ln  \left( 5 \right)
}
+m_k
\quad ,
\label{magnitudefirst}
\end{equation}
where $m_k$ is  a constant  of calibration.
This formula is useful  when we have  the  absolute/apparent
magnitude versus time as in the case  of optical LC
in SN.
The asymptotic approximation  is
\begin{equation}
m(t) \sim
1.085\alpha d\ln  \left( t \right) - 5.428\alpha
\ln  \left( t \right) + 3.257\ln  \left( t \right) +m_{{k}}+
 \frac{0.542}{ \left( {{\rm e}^{{\frac {a_{{1}}}{ \left( {t}^{-1}
 \right) ^{a_{{2}}}}}}} \right) ^2}
\quad .
\label{asymptotic}
\end{equation}

A {\it second} model for the
optical thickness assumes an exponential  law dependence
\begin{equation}
\tau_{\nu} = {\it a_{1,\nu}}\,
\left( 1-{{\rm e}^{-{\it a_{2,\nu}}\,{t}^{{\it a_{3,\nu}}}}}
\right)
\quad ,
\label{tauexp}
\end {equation}
where $a_{1,\nu}$,  $a_{2,\nu}$  and $a_{3,\nu}$
are three coefficients  which can be found
from the astronomical
data.
The  absolute/apparent magnitude version
for the {\it second} model  is
\begin{eqnarray}
m(t)  =  &  \nonumber \\
\frac
{
2.5 \ln  \left( t \right) \alpha d- 12.5 \alpha \ln  \left( t
 \right) + 7.5 \ln  \left( t \right) - 2.5 \ln  \left( 1-{{\rm e}^{-
a_{{1,\nu}} \left( 1-{{\rm e}^{-a_{{2,\nu}}{t}^{{\it a_{3,\nu}}}}} \right) }}
 \right)
}
{
\ln  \left( 2 \right) +\ln  \left( 5 \right)
}
\\
+ m_k \, .
\nonumber
\label{magnitudesecond}
\end{eqnarray}

A {\it third} model for the
optical thickness makes a comparison between
the observed intensity $I_{obs}$ and  the theoretical intensity
$I_{th}$ through the optical depth
\begin{equation}
1-e^{-\tau_{\nu}} = \frac{I_{obs}}{I_{th}}
\quad .
\end{equation}
The optical depth is
\begin{equation}
\tau_{\nu} = -\ln (1 -\frac{I_{obs}}{I_{th}})
\quad .
\end{equation}
The observed intensity as function of the time is an
astronomical quantity and the theoretical intensity can be
the mechanical luminosity or the momentary number
of radioactive isotopes.
Once the temporal behavior of  $\tau_{\nu}$    is derived
we search for the best fit as function of time.
A fit already used is the power law fit as represented
by eqn. (\ref{taupower}).
Another type of fit is the  logarithmic  polynomial approximation
of degree  M,
\begin{equation}
\tau_{\nu}(t)  =
  a_0
+ a_1 (\ln(t))
+ a_2 (\ln(t))^2
+ \cdots
+ a_M (\ln(t))^M
\quad  .
\label{polynomial}
\end{equation}
The presence of the logarithm allows  to cover
the  oscillatory behavior  of $\tau_{\nu}$ over
many decades in time.

\section{Astrophysical results}

\begin{figure}
\begin{center}
\includegraphics[width=6cm]{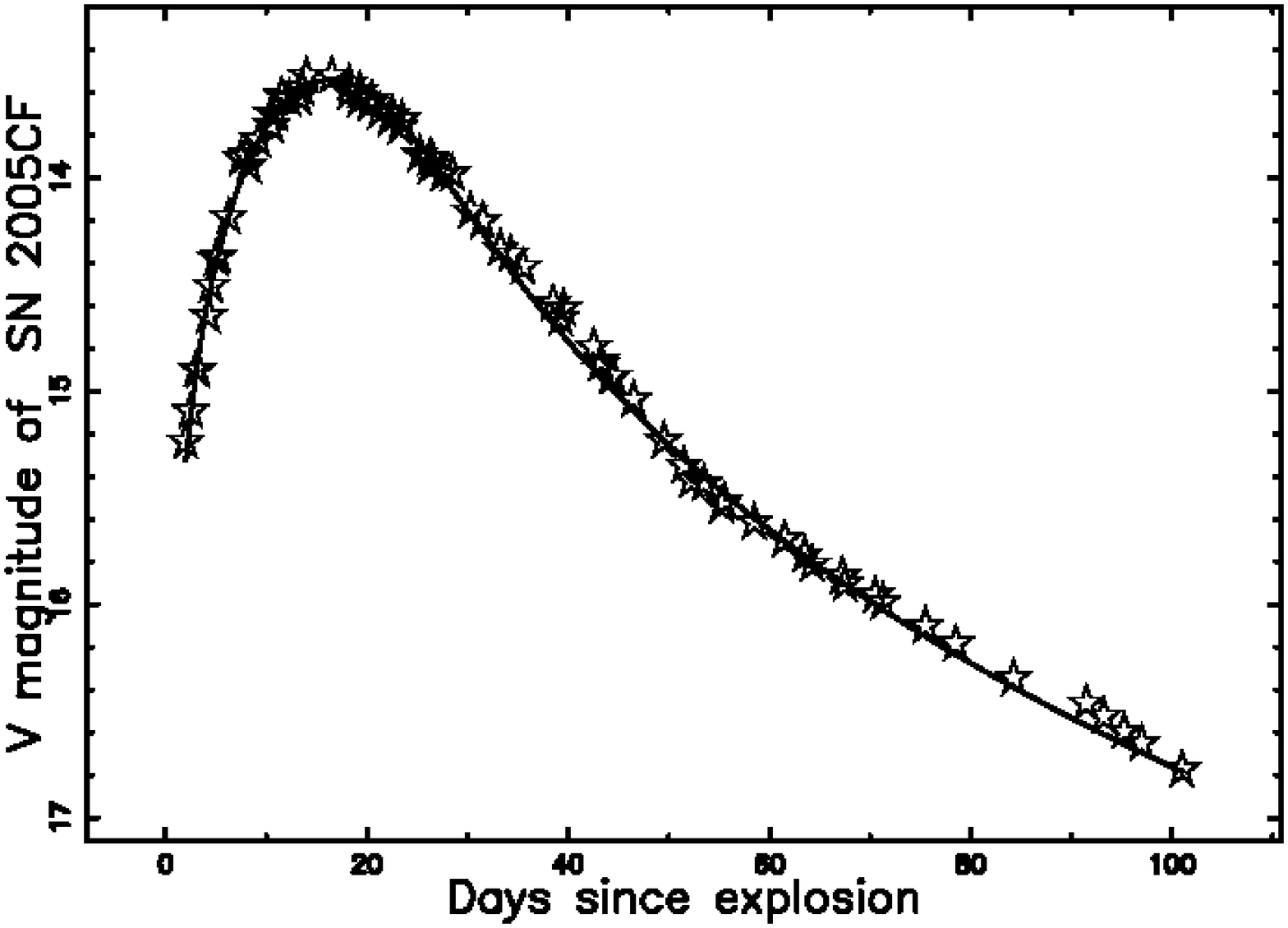}
\end {center}
\caption
{
The $V$ LC of \bsn2005 (empty stars)  and
theoretical curve as given by the first model,
see eqn. (\ref{magnitudefirst}) (full line).
}
\label{sn2005powerv}
    \end{figure}

\begin{figure}
\begin{center}
\includegraphics[width=6cm]{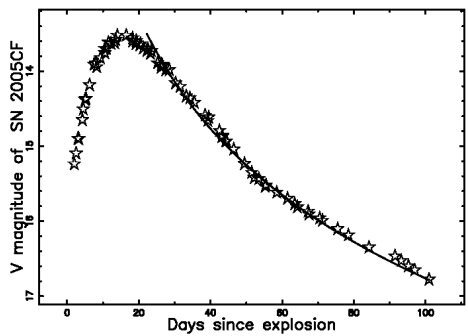}
\end {center}
\caption
{
The $V$ LC of \bsn2005 (empty stars)  and
theoretical curve as given by the asymptotic formula
(\ref{asymptotic}),
(full line).
}
\label{sn2005powerasymp}
    \end{figure}

The time is usually expressed in JD or seconds and
a subtraction of the
initial JD or seconds relative to the considered
phenomena should be done
in order to have zero at the beginning of the temporal scale.
We start by analyzing
\bsn2005 in NGC~5812 which is of type Ia, it's  distance is 29.4
Mpc and the distance moduli $\mu=32.51$,
see \cite{Pastorello2007}.
Figure \ref{sn2005powerv} reports the temporal
evolution of the $V$
visual magnitude of \bsn2005 for the power law model as well the
interpolating curve
and  Figure \ref{sn2005powerasymp} the asymptotic  approximation;
data as in Table \ref{chi2valuespower}.

The quality of the fits is measured by the
merit function
$\chi^2$
\begin{equation}
\chi^2  =
\sum_j  {(m_{th} -m_{obs})^2}
\quad ,
\nonumber
\label{chisquare}
\end{equation}
where  $m_{th}$ and  $m_{obs}$
are the theoretical and observed magnitude, respectively.

\begin{table}[ht!]
\caption
{
Numerical values
of the adopted parameters and  $\chi^2$
for the optical LC in SN in the case of
optical thickness with a power law dependence,
$\alpha=0.828$ everywhere.
}
\label{chi2valuespower}
\begin{center}
\begin{tabular}{|c|c|c|c|c|c|c|}
\hline
Name ~SN
&   band
& d
& $a_1$
& $a_2$
& $m_k$
&$\chi^2$
  \\
\hline
SN~2005cf
& V
& 3.80
& 1.9~$10^{-4}$
& 2.95
& 6.73
&0.188
  \\
SN~2005cf
& B
& 3.85
& 1.98~$10^{-4}$
& 3.2
& 7.11
& 6.7
  \\
 SN~2004A
& V
& 4.15
& 1.0~$10^{-4}$
& 4.15
& 5.33
& 12.4
  \\
 SN~2004A
& B
& 3.52
& $2.6~10^{-3}$
& 1.85
& 9.53
& 2.50
  \\
   SN~2004A
& I
& 4.33
& $2.0~10^{-5}$
& 2.65
& 3.4
& 0.35
  \\
    SN~2004A
& R
& 4.05
& $7.8~10^{-5}$
& 2.3
& 4.81
& 1.10
  \\
\hline
\end{tabular}
\end{center}
\end{table}

\begin{figure}
\begin{center}
\includegraphics[width=6cm]{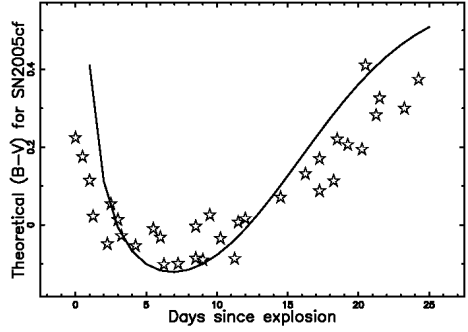}
\end {center}
\caption
{
The (B--V)  color evolution  of \bsn2005
(empty stars)  and the relative fitting
straight line (full line).
The theoretical curve is  given by the second model.
The time  is limited to the first 25 days.
\label{2005cfbvexp}
}
    \end{figure}

\begin{table}[ht!]
\caption
{
Numerical values
of the adopted parameters and  $\chi^2$
for the optical LC  in the case of
optical thickness with an exponential law dependence,
$\alpha=0.828$ everywhere.
}
\label{chi2valuesexp}
\begin{center}
\begin{tabular}{|c|c|c|c|c|c|c|c|}
\hline
Name ~SN
&   band
& d
& $a_1$
& $a_2$
& $a_3$
& $m_k$
&$\chi^2$
  \\
\hline
SN~2005cf
& V
& 3.79
& 10
& 1.75~$10^{-5}$
& 3
& 6.73
&0.202
  \\
SN~2005cf
& B
& 3.81
& 5
& 8.2~$10^{-5}$
& 3
& 7.32
& 6.45
  \\
 SN~2004A
& V
& 4.36
& 10
& 6.27~$10^{-7}$
& 3
& 4.07
& 1.70
  \\
 SN~2004A
& B
& 3.79
& 100
& $6~10^{-7}$
& 3
& 8.15
& 2.4
  \\
   SN~2004A
& I
& 4.67
& 2.5
& $1.0~10^{-6}$
& 3.0
& 1.62
& 0.52
  \\
    SN~2004A
& R
& 4.23
& 10
& $4.51~10^{-7}$
& 3
& 3.83
& 1.6
  \\
\hline
\end{tabular}
\end{center}
\end{table}
The (B--V)  color evolution  of \bsn2005 for
the exponential law
model (data as in Table \ref{chi2valuesexp})
is reported in
Figure~\ref{2005cfbvexp}.

\begin{figure}
\begin{center}
\includegraphics[width=6cm]{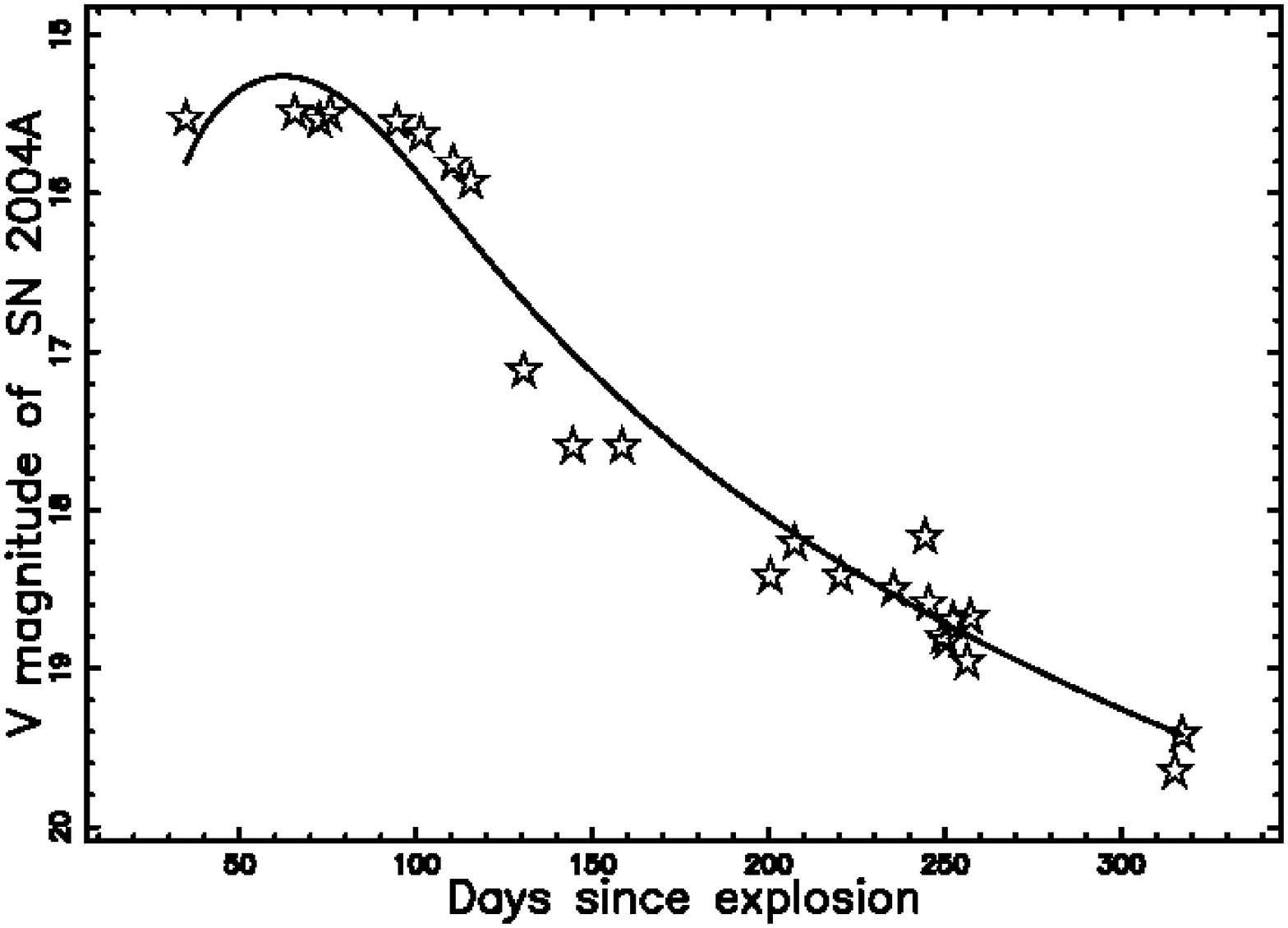}
\end {center}
\caption
{
The $V$ LC of \csn2004 (empty stars)  and
theoretical curve as given by the second model,
see eqn. (\ref{magnitudesecond})
(full line).
}
\label{2004a_v_exp}
    \end{figure}

The SNs of type IIp are characterized  by a flat LC for a long
period of time, i.e. 100 days.
We therefore  analyzed \csn2004 in
NGC~6207 , which is of type IIp, the distance is  25.6 Mpc
and the
distance moduli $\mu=31.99$, see \cite{Hendry2006,Tsvetkov2008}.
Figure \ref{2004a_v_exp}  reports the temporal evolution
of the $V$
visual magnitude of \csn2004 for the
exponential law model as well
the interpolating curve,
data as in Table \ref{chi2valuesexp}.

\begin{figure}
\begin{center}
\includegraphics[width=6cm]{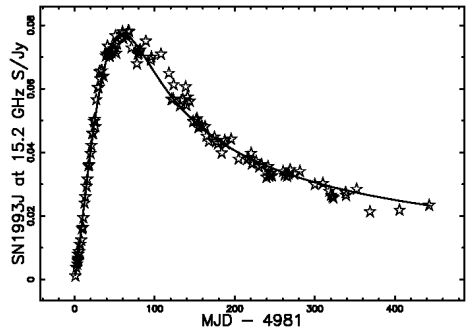}
\end {center}
\caption
{
The radio flux density   of \asn1993
at 15.2 GHz (empty stars)  and
theoretical curve as given by the first model,
see eqn. (\ref{fluxfirst})
(full line).
}
\label{sn1993jradio}
    \end{figure}

We now apply  the developed theory to model the
radio flux density
of \asn1993 at 15.2 GHz,
see \cite{Pooley1993,Ho1999}, with data
available at \newline
http://www.mrao.cam.ac.uk/~dag/sn1993j.html .
In this radio-case  we plot the flux version
of the first model as  given by  eqn.(\ref{fluxfirst}),
see Figure~\ref{sn1993jradio}
and Table \ref{chi2valuesradiogamma}.

\begin{table}[ht!]
\caption
{
Numerical values
of the adopted parameters and  $\chi^2$
for the radio LC in \asn1993 and gamma LC in
\grb  in the case of
optical thickness with a power law dependence,
$\alpha=0.828$ everywhere.
}
\label{chi2valuesradiogamma}
\begin{center}
\begin{tabular}{|c|c|c|c|c|c|c|}
\hline
Name
&   band
& d
& $a_1$
& $a_2$
& $F_{\nu,0}$
&$\chi^2$
  \\
\hline
SN\,1993J
& 15.2 GHz
& 2.22
& 8.5~$10^{-4}$
& 1.86
& 1.64
&  7.9~$10^{-4}$
\\
GRB 050814
& 0.2-10 kev
& 2.79
&  0.026
& 1.259
& $8.38 10^{-08}$
& 4.9~$10^{-18}$
\\
\hline
\end{tabular}
\end{center}
\end{table}

\begin{figure}
\begin{center}
\includegraphics[width=6cm]{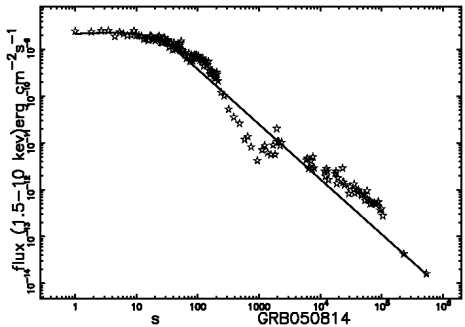}
\end {center}
\caption
{
The XRT flux  of \grb
at 0.2-10 kev  (empty stars)  and
theoretical curve as given by the first model,
see eqn. (\ref{fluxfirst})
(full line).
}
\label{grb050814}
    \end{figure}

The theory is now applied to
\grb
at 0.3-10 kev  in the time interval
$10^{-5}-3$ days,
see \cite{Jakobsson2006}
with data available at
\newline
http://www.swift.ac.uk/xrt$\_$curves/00150314/.
Figure
\ref{grb050814} reports  the  LC, in this case the flux,
as function of the elapsed
time since  Burst Alert Telescope (BAT) trigger and Table
\ref{chi2valuesradiogamma} reports the involved parameters.

\section{Relativistic model}

The density, $\rho$,
of the ISM at a distance $r$ from the
SN is here  modeled  by  a
Lane--Emden ($n=5$) profile
\begin{equation}
\rho(r;b) =\rho_c {\frac {1}{{(1+ \frac{{r}^{2}}{3b^2})^{5/2}}} }
\label{densita5b}
\quad ,
\end{equation}
where b represents the scale.
The relativistic conservation of momentum
for the thin layer approximation
in presence of a  the  Lane--Emden ($n=5$) profile
is given  by  the following   differential
equation
\begin{equation}
\frac{
4\,{b}^{3} ( r ( t )  ) ^{3}\rho\,\pi \,\sqrt {3
}{\frac {d}{dt}}r ( t )
}
{
( 3\,{b}^{2}+ ( r ( t )  ) ^{2} ) ^{
3/2}c\sqrt {-{\frac { ( {\frac {d}{dt}}r ( t )
 ) ^{2}}{{c}^{2}}}+1}
}
= \frac {
4\,{b}^{3}{r_{{0}}}^{3}\rho\,\pi \,\sqrt {3}\beta_{{0}}
}
{
( 3\,{b}^{2}+{r_{{0}}}^{2} ) ^{3/2}\sqrt {-{\beta_{{0}}}^{
2}+1}
}
\quad ,
\label{eqndiffrel}
\end{equation}
where $r_0$  is the initial radius of the advancing sphere,
$v_0$
is the initial velocity at $r_0$,
c is  the light  velocity  and
$\beta_0=\frac{v_0}{c}$.
The relativistic transfer  of  energy
through a surface, $A$, is
\begin{equation}
L_{m,r}  = A \gamma^2 ( \rho c^2 +p ) v
\quad ,
\end{equation}
where $p$  is the pressure, for sake of simplicity we take p=0,
 and
the Lorentz factor  $\gamma$ is
\begin{equation}
\gamma = \frac{1} {\sqrt{1-\beta^2} dt}
\quad ,
\end{equation}
see eqn. A31  in \cite{deyoung} or eqn. (43.44) in
\cite{Mihalas2013}.

\begin{figure*}
\begin{center}
\includegraphics[width=7cm]{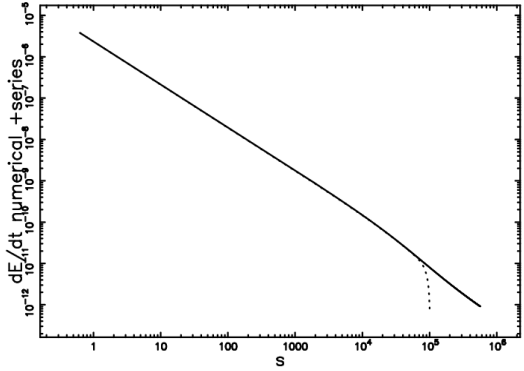}
\end {center}
\caption
{
Numerical $L_{m,r}$
computed according to  Eq. (\ref{luminosityrel}) (full line)
and series solution as
given by Eq. (\ref{luminosityseries}) (dotted line).
Data as in Table \ref{datafitrel}.
}
\label{series_numer_rel}
    \end{figure*}

In the case of a spherical cold expansion
\begin{equation}
L_{m,r} = 4 \pi r(t)^2 \frac{1}{1-\beta(t)^2} \rho(t) c^3 \beta(t)
\quad .
\end{equation}
We now assume the  following  power law behavior  for the
density in the advancing thin layer
\begin{equation}
\rho(t) = \rho_0 (\frac{t_0} {t})^d
\quad ,
\end{equation}
and we obtain
\begin{equation}
L_{m,r} =
4 \pi r(t)^2
\frac{1}{1-\beta(t)^2}
\rho_0 (\frac{t_0} {t})^d
c^3 \beta(t)
\quad .
\label{luminosityrel}
\end{equation}
We can now  derive $L_{m,r}$ in two ways: (i) from a numerical
evaluation of r(t) and  v(t),
(ii) from a Taylor series  of
$L_{m,r}(t)$ of the type
\begin{equation}
L_{m,r}(t) =
\sum _{n=0}^{3}a_{{n}}{(t-t_0)}^{n}
\quad .
\label{luminosityseries}
\end{equation}
The coefficients are
\begin{eqnarray}
a_0=&  \frac{
-4\,\pi \,{r_{{0}}}^{2} \left( {\frac {t_{{0}}}{t}} \right) ^{d}{c}^{3
}\beta_{{0}}
}
{
{\beta_{{0}}}^{2}-1
}
  \nonumber \\
a_1 =&  \frac{
4\,\pi \,r_{{0}}{c}^{4}{\beta_{{0}}}^{2} \left( {\frac {t_{{0}}}{t}}
 \right) ^{d} \left( 9\,{b}^{2}{\beta_{{0}}}^{2}+3\,{b}^{2}-2\,{r_{{0}
}}^{2} \right)
}
{
 \left( 3\,{b}^{2}+{r_{{0}}}^{2} \right)  \left( {\beta_{{0}}}^{2}-1
 \right)
}
  \label{acoeff} \\
a_2 =&  \frac{
2\,{\beta_{{0}}}^{3}{c}^{5} \left( {\frac {t_{{0}}}{t}} \right) ^{d}
\pi \, \left( 162\,{b}^{4}{\beta_{{0}}}^{4}-297\,{b}^{4}{\beta_{{0}}}^
{2}-9\,{b}^{2}{\beta_{{0}}}^{2}{r_{{0}}}^{2}-45\,{b}^{4}+15\,{b}^{2}{r
_{{0}}}^{2}-2\,{r_{{0}}}^{4} \right)
}
{
 \left( 3\,{b}^{2}+{r_{{0}}}^{2} \right) ^{2} \left( {\beta_{{0}}}^{2}
-1 \right)
}
  \nonumber \\
 a_3 =&  \frac{
 18\, \left( 270\,{b}^{2}{\beta_{{0}}}^{6}-675\,{b}^{2}{\beta_{{0}}}^{4
}-33\,{\beta_{{0}}}^{4}{r_{{0}}}^{2}+480\,{b}^{2}{\beta_{{0}}}^{2}+62
\,{\beta_{{0}}}^{2}{r_{{0}}}^{2}+45\,{b}^{2}-5\,{r_{{0}}}^{2} \right)
{\beta_{{0}}}^{4}{b}^{4}{c}^{6}\pi \, \left( {\frac {t_{{0}}}{t}}
 \right) ^{d}
}
{
r_{{0}} \left( 27\,{b}^{6}+27\,{b}^{4}{r_{{0}}}^{2}+9\,{b}^{2}{r_{{0}}
}^{4}+{r_{{0}}}^{6} \right)  \left( {\beta_{{0}}}^{2}-1 \right)
}
 \nonumber
\quad .
\end{eqnarray}
Figure \ref{series_numer_rel} compares
the numerical  solution for the luminosity and
the series expansion for  the
luminosity about the  ordinary point $t_0$.

\begin{table}
\caption
{
Numerical values of the parameters
used in  relativistic solutions.
}
 \label{datafitrel}
 \[
 \begin{array}{c}
 \hline
 \hline
 \noalign{\smallskip}
 parameters      \\
  t_0=2^{-8}~\mathrm{yr}~or~t_0=0.63~\mathrm{s}~;
  r_0=0.00195~\mathrm{pc} ~;
~\beta_0=0.833 ~;
~ b=0.004~ \mathrm{pc}
\\
\noalign{\smallskip} \hline
 \end{array}
 \]
 \end {table}

\begin{figure}
\begin{center}
\includegraphics[width=6cm]{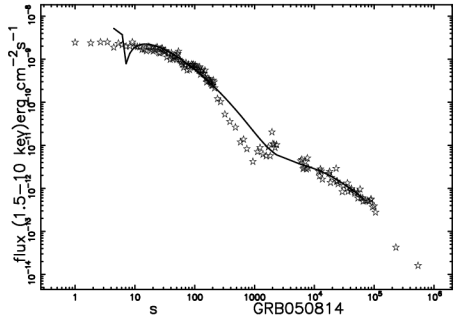}
\end {center}
\caption
{
The XRT flux  of \grb
at 0.2-10 kev  (empty stars)  and
theoretical curve as given by the relativistic numerical model,
see eqn. (\ref{fluxrel})
(full line).
}
\label{grb050814_rel}
    \end{figure}

The flux  at  frequency  $\nu$ and  distance  $D$
is
\begin{equation}
F_{\nu,r} = \frac
{\epsilon_{\nu} L_{m,r} }
{4 \pi D^2}
\quad .
\end{equation}
The flux corrected for absorption in the relativistic case
is
\begin{equation}
F_{\nu,c,r}=
\frac
{\epsilon_{\nu}  L_{m,r}}
{4 \pi D^2}
(1-e^{-\tau_{\nu}})
\quad .
\label{fluxrel}
\end{equation}
As a behavior for  ${\tau_{\nu}}$ as function of time
we select a logarithmic polynomial approximation,
see (eqn.\ref{polynomial}),
of degree 9 and
Figure \ref{grb050814_rel} reports  the
flux of the relativistic
LC as function of the elapsed
time since  Burst Alert Telescope (BAT) trigger.

\section{The radioactive model}
Here we consider the decay  of a radioactive  isotope,
a radioactive chain  and
the Arnett's rule for the bolometric luminosity.

\subsection{Decay of one element}

The decay of a radioactive  isotope is  modeled by the
following equation
\begin{equation}
-dN = \frac{N}{\tau_n} dt
\quad ,
\end {equation}
where $\tau_n$ is a constant  and the negative sign
indicates that $dN$ is a reduction in the number
of nuclei , see  \cite{Yang2010}.
The integration of this differential equation
of the first order in which the variables can be separated
gives:
\begin{equation}
N(t) = N_0 e^{-\frac{t}{\tau_n}}
\label{ntradioactive}
\quad ,
\end{equation}
where $N_0$ is the number of nuclei at  $t=0$.
The half life is  $T_{1/2}= ln(2) \; \tau_n$.
The absolute magnitude version  of the previous formula
is
\begin{equation}
M = - C \; Log_{10} (N(t)) = {-\frac{t}{\tau_n}} +k
\quad ,
\label{magnituderadioactive}
\end{equation}
where $M$ is the absolute luminosity, $C$ and $k$ are two constants.
This means that we are waiting  for a straight line
for the absolute magnitude versus time relationship.
At the same time the observational  fact
that  the spectral index in the radio
varies  considerably  but  becomes constant, $\beta \approx  -
0.7$, after  $\approx$ 700 days, see Figure 8 in
\cite{MartiVidalb2011}, asks the  absorption.

\subsection{Radioactive chains}

\begin{figure}
\begin{center}
\includegraphics[width=6cm]{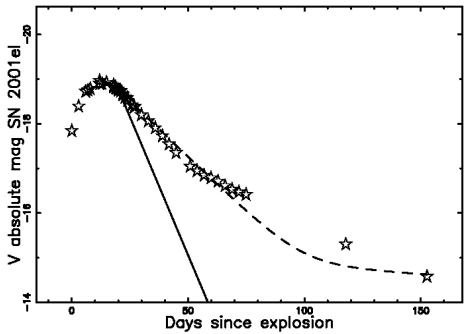}
\end {center}
\caption
{
The $V$
LC of \sn2001el (empty stars)
in absolute magnitude,
the theoretical curve as given
by equation (\ref{magnituderadioactive})
when
the radioactive decay of the  isotope $^{56}$Ni
($\tau_n $ = 8.757 d  or  $T_{1/2}$ =6.07 d , $k$=-18.65)
was considered
(full line),
and
the theoretical curve, (dashed line),   connected with
the decay of two species as represented
by eq.(\ref{magnitudechain}) ($\tau_1=8.767 d, \tau_2=111.477 d,N_{0,1}=1$,
$C_1$=1.9995, $C_2$=1.0005
and $k$=-18.4).
}
\label{chain}
    \end{figure}

The  isotope $^{56}$Ni is unstable  and decays
($\tau_1 $ = 8.757 d,  $T_{1/2}$ =6.07 d)
into $^{56}$Co emitting gamma photons.
The  isotope $^{56}$Co is unstable and decays
($\tau_2 $ = 111.47 d,  $T_{1/2}$ = 77.27 d) into $^{56}$Fe
through electron capture
and $\beta$-decay.
The decay rates of of the two species, species 1 is $^{56}$Ni and
species 2 is $^{56}$Co,
 is  modeled by the
following equations
\begin{subequations}
\begin{align}
{\frac {\rm d}{{\rm d}t}}N_{{1}} \left( t \right)&=-{\frac {N_{{1}}
 \left( t \right) }{\tau_{{1}}}}\\
  {\frac {\rm d}{{\rm d}t}}N_{{2}} \left( t \right)&={\frac {N_{{1}}
 \left( t \right) }{\tau_{{1}}}}-{\frac {N_{{2}} \left( t \right) }{
\tau_{{2}}}}.
\end{align}
\end{subequations}
The two solutions obtained inserting as initial conditions
$N_1(0)=N_{0,1}$ and  $N_2(0) =0$ are
\begin{subequations}
\begin{align}
N_1&=N_{{0,1}}{{\rm e}^{-{\frac {t}{\tau_{{1}}}}}}
\label{N1}
\\
N_2&= \left( {\frac {\tau_{{2}}N_{{0,1}}}{\tau_{{1}}-\tau_{{2}}}{{\rm e}^{-
{\frac {t}{\tau_{{1}}}}+{\frac {t}{\tau_{{2}}}}}}}-{\frac {\tau_{{2}}N
_{{0,1}}}{\tau_{{1}}-\tau_{{2}}}} \right) {{\rm e}^{-{\frac {t}{\tau_{
{2}}}}}}.
\label{N2}
\end{align}
\end{subequations}
The sum of the two species, N(t), is
according to formula (8.5) in \cite{Rust2010}
\begin{equation}
N(t) = C_1 \, N_1(t) + C_2 \,N_2(t)
\quad ,
\label{sumchain}
\end{equation}
where  $C_1$ and $C_2$  are two adjustable parameters.
This  linear sum
is associated with the LC in SNs assuming that the $\gamma$-rays
are thermalized in the ejecta and emerge in the various bands.
The logarithmic form,M(t) , is associated with  the magnitude evolution
\begin{equation}
\label{magnitudechain}
M = \frac{MN}{MD}
\end{equation}
where
\begin{eqnarray}
MN = k\ln   ( 2  ) +k\ln   ( 5  )
-\ln   ( N_{{0,1}}  )   \nonumber \\
-\ln   ( {\frac {1
}{\tau_{{1}}-\tau_{{2}}}  ( C_{{2}}\tau_{{2}}{{\rm e}^{-{\frac {t
}{\tau_{{2}}}}}}{{\rm e}^{{\frac {t  ( \tau_{{1}}-\tau_{{2}}
  ) }{\tau_{{1}}\tau_{{2}}}}}}+C_{{1}}{{\rm e}^{-{\frac {t}{\tau_
{{1}}}}}}\tau_{{1}}-C_{{1}}{{\rm e}^{-{\frac {t}{\tau_{{1}}}}}}\tau_{{
2}}-C_{{2}}\tau_{{2}}{{\rm e}^{-{\frac {t}{\tau_{{2}}}}}}  ) }
  )
\quad ,
\end{eqnarray}
where k is a constant
and
\begin{equation}
Md = \ln   ( 2  ) +\ln   ( 5  )
\quad .
\end{equation}
We  plot  the decay
of the LC of  \sn2001el , which is of type Ia,
adopting a  distance
modulus of 31.65 mag,
see  \cite{Krisciunas2003},
the  nuclear  decay
which  according to
equation (\ref{magnituderadioactive})
is a straight line,
and the theoretical curve
of the two species
as  represented by equation \ref{magnitudechain},
see  Figure \ref{chain}.

\subsection{Bolometric luminosity}

\begin{figure}
\begin{center}
\includegraphics[width=6cm]{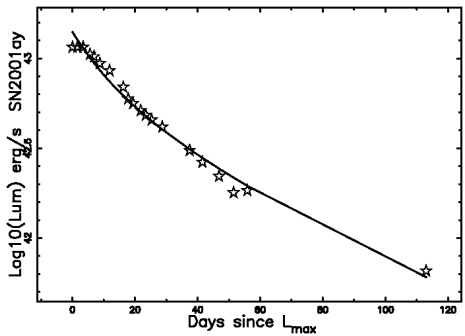}
\end {center}
\caption {
The bolometric LC
of  \dsn2001ay (empty stars)
and
the theoretical curve as given
by eqn. (\ref{bolocorrected})
(full line).
The parameters are $t_R$=22 days , $\alpha=1$
and  the comparison method with a power law fit
gives $a_1= 105.88$ and $a_2=-1.335$.
The astronomical data are extracted
from Figure 18 in \cite{Krisciunas2011}
by the author.
}
\label{2001aybolo}
    \end{figure}

The bolometric  LC after \cite{Arnett1982,Arnett1985}
has  been associated with the combined
radioactive decays of the  isotopes $^{56}$Ni
and $^{56}$Co.
A formula of practical use is
given by
\begin{equation}
L(t_R)  = \alpha \left ( (6.45 \times 10^{43}) e^{-t_R/8.8} +
                           (1.45 \times 10^{43}) e^{-t_R/111.3} \right)
\frac{erg}{s}
\quad ,
\end{equation}
where $t_R$  is the elapsed time from the explosion to the
maximum of the LC and $\alpha$ is $\approx$ 1,
see formula (2) in \cite{Krisciunas2011}.
The previous formula represents the optically thin case.
According to the comparison method developed in Section
\ref{sectransition} a corrected bolometric luminosity
for the absorption, $L_c(t_R)$, is
\begin{equation}
L_c(t_R)=L(t_R)\times \left( 1-{{\rm e}^{-
a_{{1}}{t}^{a_{{2}}}}} \right)
\quad  ,
\label{bolocorrected}
\end{equation}
and  Figure~\ref{2001aybolo} reports the comparison between
observed and theoretical bolometric luminosity.

\section{Conclusions}

{\it Classical and relativistic flux of energy:}
The classical flux of kinetic energy  can be easily parametrized
in the case of a radius-time relationship represented
by  a power law, see eqn.(\ref{lmalpha}).
Conversely is more complex to derive the relativistic flux of
kinetic energy which requires a relativistic law of motion.
In the framework of
Lane--Emden ($n=5$) profile as given by eqn.(\ref{densita5b})
and momentum conservation in a thin layer we can deduce an
analytical solution for the relativistic flux of energy in terms
of a Taylor series, see the four coefficients in (\ref{acoeff}).

{\it Light curve:}
Assuming a linear relationship
between  the luminosity in the various  astronomical
bands and the classical or relativistic
flux  of mechanical kinetic energy
we can  easily  deduce a theoretical  time dependence
for the LC,
see  classical  eqn.(\ref{flux_classical})
or relativistic eqn.(\ref{luminosityseries}).
This theoretical dependence  is not enough
and the concept of optical depth
should be introduced.
Among the infinite  relationships  for  optical depth
as function of time we selected  a power law
dependence, see eqn.(\ref{taupower}),
an exponential
behavior,  see eqn.(\ref{tauexp}),
or a logarithmic  polynomial approximation,
see eqn.(\ref{polynomial}).

{\it Nuclear Decay:}
The  LC of a SN is often model
by the decay of the radioactive  isotope
$^{56}$Ni, but in order to follow the LC with time
we  should  introduce a  radioactive chain,
see eqn. (\ref{sumchain}).
Further on some classical approach to the
bolometric luminosity must be corrected
for the optical depth, see eqn.(\ref{bolocorrected}).

{\it Comparison with astronomical data:}
The framework  of conversion   of
the classical flux  of mechanical kinetic energy
into the various  optical bands coupled
with a time dependence  for the optical depth
allows to simulate the  various morphologies of the
LC: for a type Ia  we chosen  \bsn2005,
see Figure \ref{sn2005powerv} for V band  and
Figure  \ref{2005cfbvexp} for (B-V) color.
The enigmatic behavior  of type IIp  SNs,
here represented by  \csn2004,
can also
be modeled, see Figure \ref{2004a_v_exp} for the V band
and Table \ref{chi2valuespower}  for B, I and R bands.
The opposite sides  of the electro-magnetic spectrum
can  also be simulated:  for  the radio band
of  \asn1993 see Figure~\ref{sn1993jradio}
and for the gamma/X spectrum  of \grb\,
  see Figure \ref{grb050814}.
More complex is the derivation of the relativistic
flux of energy here parametrized by a series
expansion. The coupling of the previous series
with a logarithmic  polynomial approximation
allows to model fine details such as the
oscillation in LC visible at $\approx$ 1000 s in
\grb, see  Figure \ref{grb050814_rel}.
All the fits here presented report the $\chi^2$,
see Tables \ref{chi2valuespower} and \ref{chi2valuesexp}.
This means that other types of functions
for the optical depth versus time have a reference for comparison.

\end{document}